\pdfoutput=1

\documentclass[11pt]{article}

\usepackage[]{acl}
\usepackage{times}
\usepackage{booktabs}

\usepackage{times}
\usepackage{latexsym}
\usepackage[T1]{fontenc}
\usepackage[utf8]{inputenc}
\usepackage{microtype}
\usepackage{inconsolata}
\usepackage{xspace}
\usepackage{amsmath}
\usepackage{amssymb}
\usepackage{mathtools}
\usepackage{tcolorbox}
\usepackage{pgfplots}
\pgfplotsset{compat=1.7}
\usepackage{caption}
\usepackage{subcaption}
\usepackage[hang,flushmargin]{footmisc}
\usepackage{multirow}
\usepackage{enumitem}
\usepackage{arydshln}
\usepackage{color}
\usepackage{tcolorbox}
\usepackage{CJK}
\usepackage{adjustbox}
\usepackage{float} 
\usepackage[table]{xcolor}
\usepackage[T1]{fontenc}
\usepackage[utf8]{inputenc}
\usepackage{microtype}
\usepackage{inconsolata}
\usepackage{placeins}
\usepackage[title]{appendix}

\newcommand{\rankVicuna}{Rank\-Vicuna\xspace}
\newcommand{\rankZephyr}{Rank\-Zephyr\xspace}

\newcommand{\ScoreFiD}{LiT5-Score\xspace}
\newcommand{\RankFiD}{LiT5-Distill\xspace}
\newcommand{\RankFiDbase}{LiT5-Distill\textsubscript{base}\xspace}
\newcommand{\RankFiDlarge}{LiT5-Distill\textsubscript{large}\xspace}
\newcommand{\RankFiDxl}{LiT5-Distill\textsubscript{XL}\xspace}
\newcommand{\ScoreFiDbase}{LiT5-Score\textsubscript{base}\xspace}
\newcommand{\ScoreFiDlarge}{LiT5-Score\textsubscript{large}\xspace}
\newcommand{\ScoreFiDxl}{LiT5-Score\textsubscript{XL}\xspace}

\newcommand{\gptthreefive}{GPT\textsubscript{3.5}\xspace}
\newcommand{\gptfour}{GPT\textsubscript{4}\xspace}
\newcommand{\rankgptthreefive}{RankGPT\textsubscript{3.5}\xspace}
\newcommand{\rankgptfour}{RankGPT\textsubscript{4}\xspace}
\newcommand{\ada}{ADA\textsubscript{2}\xspace}

\newcommand{\rankllama}{Rank\-LLaMA\xspace}
\newcommand{\repllama}{Rep\-LLaMA\xspace}

\title{Scaling Down, LiTting Up: Efficient Zero-Shot Listwise Reranking with Seq2seq Encoder--Decoder Models}

\author{Manveer Singh Tamber, Ronak Pradeep, Jimmy Lin \\[1ex]
David R. Cheriton School of Computer Science,\\ University of Waterloo, Canada \\[1ex]
\texttt{\{mtamber, rpradeep, jimmylin\}@uwaterloo.ca}}
\begin{document}
\maketitle

\begin{abstract}
Recent work in zero-shot listwise reranking using LLMs has achieved state-of-the-art results.
However, these methods are not without drawbacks.
The proposed methods rely on large LLMs with billions of parameters and limited context sizes.
This paper introduces \RankFiD and \ScoreFiD, two methods for efficient zero-shot listwise reranking, leveraging T5 sequence-to-sequence encoder--decoder models.
Our approaches demonstrate competitive reranking effectiveness compared to recent state-of-the-art LLM rerankers with substantially smaller models.
Through \ScoreFiD, we also explore the use of cross-attention to calculate relevance scores to perform reranking, eliminating the reliance on external passage relevance labels for training.
We present a range of models from 220M parameters to 3B parameters, all with strong reranking results, challenging the necessity of large-scale models for effective zero-shot reranking and opening avenues for more efficient listwise reranking solutions.
We provide code and scripts to reproduce our results at \url{https://github.com/castorini/LiT5}.
\end{abstract}

\section{Introduction}

Listwise reranking using LLMs has seen success in recent work and has attained state-of-the-art results for zero-shot reranking~\cite{RankGPT, pradeep2023rankvicuna, pradeep2023rankzephyr, ma2023zeroshot}.
These approaches leverage the extensive capabilities of LLMs to take a query and a list of passages and output a reordered ranked list.
However, these listwise rerankers rely on large LLMs with billions of parameters and limited context sizes.
This reliance on large-scale models introduces challenges in terms of computational demands.

Additionally, existing methods for training LLMs to perform listwise reranking~\cite{pradeep2023rankvicuna, pradeep2023rankzephyr, zhang2023rankwithoutgpt} rely on training using passage relevance labels from either human judgements or in the form of rankings generated by a teacher model.
This strategy contrasts with leveraging just the intrinsic ability of LLMs to comprehend and process information for ranking purposes, which has seen success in recent work~\cite{RankGPT, qin2023large, ma2023zeroshot}.

This paper introduces two List-in-T5 (LiT5) models for listwise reranking using the T5 model~\cite{t5}: \RankFiD and \ScoreFiD, both taking inspiration from the Fusion-in-Decoder (FiD)~\cite{FiD} architecture to perform efficient listwise reranking with encoder--decoder models.

Our investigation was steered by the following research questions:

\begin{itemize}[leftmargin=*]

\item Can sequence-to-sequence encoder--decoder models be adapted for listwise reranking?

\item Can the reranking effectiveness of much larger models be distilled into our smaller sequence-to-sequence encoder--decoder reranking models?

\item Can cross-attention scores from a FiD model trained for question-answering be used to rerank passages in a zero-shot setting?

\item How much improvement is seen in zero-shot reranking effectiveness as the models are scaled up in parameters?

\end{itemize}

\smallskip
\noindent
Our work demonstrates that encoder--decoder-based methods work well for listwise reranking and can attain high reranking effectiveness competitive with recent work on zero-shot reranking using LLMs, despite the dramatically smaller model sizes.
We successfully distill reranking effectiveness from much larger models to models orders of magnitude smaller.
We also show that reranking using relevance scores from cross-attention is an effective strategy that capitalizes on the ability of FiD models to interpret and respond to queries based on context passages, thereby eliminating the reliance on external passage relevance labels to train listwise rerankers.
Additionally, our findings highlight an interesting dynamic in model scaling:\
While larger models do exhibit improvements in reranking effectiveness, we demonstrate that much smaller models can still deliver competitive results.

\section{Background and Related Work}

\subsection{Fusion-in-Decoder (FiD)}

The Fusion-in-Decoder (FiD)~\cite{FiD} model is a retrieval-augmented language model that has achieved state-of-the-art results on multiple knowledge-intensive tasks such as open-domain question-answering~\cite{atlasqa}.
The FiD model modified the T5 Encoder--Decoder model to take multiple passages, encode the passages separately, and then perform attention over the concatenation of all retrieved passages in the decoder.
In the original and subsequent FiD work, the FiD model read context passages retrieved from a corpus of Wikipedia passages to answer questions~\cite{FiD, DKRR, tamber2023pre, GAR, atlasqa}.

The FiD model is a well-suited architecture for retrieval-augmented generation because it can read many retrieved passages efficiently.
Since the FiD model encodes each passage separately, the computation scales linearly with the number of passages in the encoder.
After the encoder, the decoder also has computation complexity linear in the number of passages.
This is because the decoder performs attention over all the encoded representations of the input passages to produce an answer with computation complexity linear in the number of tokens across all passages~\cite{de2022fido}.
The encoder reads each passage separately with the same shared parameters and the decoder consults multiple passages and uses information from these passages to produce an answer.
With decoder-only LLMs, the computation scales quadratically with the number of passages considered unless some sparse attention method is used~\cite{child2019generating}.

\subsection{Distilling from RankGPT}

RankGPT~\cite{RankGPT} showed that both \gptthreefive and \gptfour are strong zero-shot listwise rerankers with \rankgptfour achieving state-of-the-art results for zero-shot reranking.
The work further showed that the reranking effectiveness of RankGPT could be distilled into smaller pointwise models.
Subsequent work showed that the reranking effectiveness of both \rankgptthreefive and \rankgptfour could be distilled into smaller open-source LLMs for listwise reranking that could, in some cases, surpass their respective teachers in reranking~\cite{ pradeep2023rankvicuna, pradeep2023rankzephyr}.
The distillation process involved leveraging instruction fine-tuning to train student models to generate the same ranking orderings as the teacher GPT models.

In our \RankFiD method we attempt to do just this.
We distill ranking orderings from teacher \rankgptthreefive and \rankgptfour models to perform reranking with much smaller student encoder--decoder models.

\begin{figure*}[ht]
\centering
\includegraphics[width=0.8\textwidth]{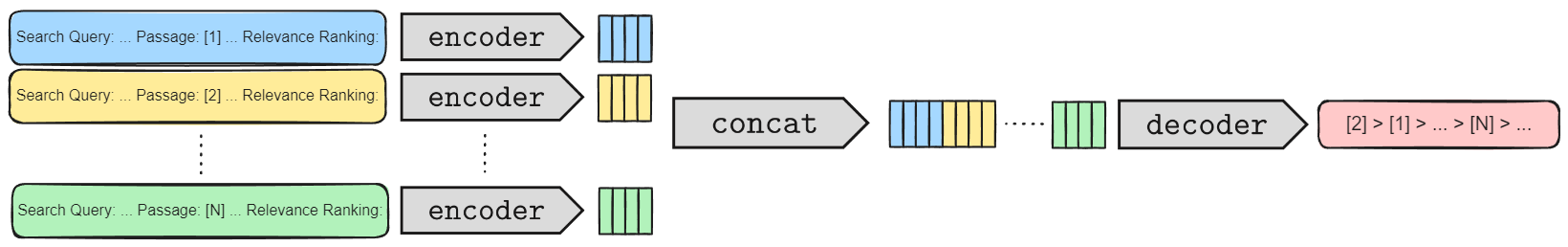}

\caption{\RankFiD architecture. Each query--passage pair is encoded separately. Then, the decoder reads over the concatenated representations to generate a ranking of the form [] > [] > ... > [], eg., [2] > [1] > ... > [N].}

\label{fig:rankfid}

\end{figure*}

\begin{figure*}[ht]
\centering
\includegraphics[width=0.8\textwidth]{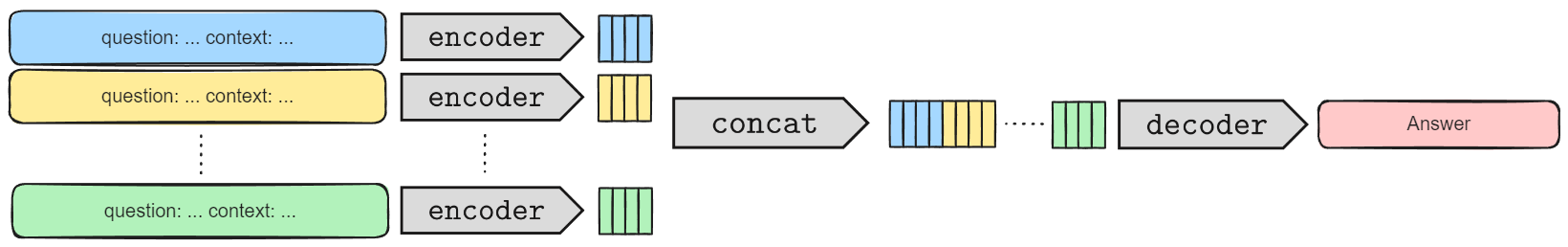}

\caption{\ScoreFiD architecture. Each query--passage (or question--context) pair is encoded separately. Then, the decoder reads over the concatenated representations to generate an answer to the query.}

\label{fig:scorefid}

\end{figure*}

\subsection{Relevance Scores with Cross-Attention}

The Distilling Knowledge from Reader to Retriever work by~\citet{DKRR} introduced a dense retriever that leveraged cross-attention scores from the FiD model to obtain relevance labels to train the retriever.
Given some context passages and a question, a relevance label calculated using cross-attention scores from the FiD model can be extracted for each context passage with respect to the question as the model generates an answer, revealing a measurement of how important the passage is to generating the answer.
The authors argued that the more the tokens in the text passage are attended to, the more relevant the passage is to answering the question.
Relevance scores are obtained through aggregating attention scores by taking the average of the attention scores over all the tokens in the passage and all the layers and heads of the decoder.

The calculation of the relevance scores for each document was improved in subsequent work~\cite{atlasqa} where in the encoder--decoder attention mechanism defined by $y = \sum_{n=1}^{N} \alpha_n v_n$ for output $y$, the relevance score is taken not just as the attention score $\alpha_n $ for token $n$, but also takes the norm of the value vector of token $n$ so that the quantity $\alpha_n ||v_n||_2$ measures the relevance of some token $n$.
This quantity is then averaged over all attention heads, layers, and tokens to calculate a relevance score for each passage.
This quantity was found to be a better measure to use to measure passage relevance~\cite{atlasqa}.

Our \ScoreFiD method leverages relevance scores calculated from the cross-attention scores to perform reranking in a zero-shot setting.

\section{Methods}

\subsection{\RankFiD}

In the \RankFiD method, following other work~\cite{pradeep2023rankvicuna, pradeep2023rankzephyr}, we use RankGPT as a teacher model to distill down ranking orderings to our student \RankFiD.

Figure~\ref{fig:rankfid} shows the setup of \RankFiD.
Similar to FiD, the model processes each passage separately alongside the query.
The input prompt for each query--passage pair starts with \texttt{Search Query:} followed by the query and then \texttt{Passage:} followed by a numerical identifier in square brackets, e.g., [1], [2], and then the passage text.
Finally, the input prompt ends with \texttt{Relevance Ranking:} to prompt the model to generate a ranking.
Then the decoder has the task of reading over the concatenated representations of each input for each passage to generate a ranking in the same output format as RankGPT, \rankVicuna, and \rankZephyr.
That is, an ordering of the passage identifiers in decreasing order of relevance in the form [] > [] > ... > [], e.g., [2] > [1] > ... > [20].

To reorder 100 passages, we use the same sliding window strategy as RankGPT, \rankVicuna, and \rankZephyr---with the same window size of 20 and stride of 10---to distill ranking orderings of up to 20 passages from RankGPT.
We note that these previous models use a small window size due to their limited context sizes and that our models can rerank more passages in a single pass, as is the case with \ScoreFiD.
Additionally, to keep results comparable, we only perform one sliding window pass, although multiple passes may help reranking effectiveness as shown with \rankZephyr~\cite{pradeep2023rankzephyr}.

\subsection{\ScoreFiD}
In the \ScoreFiD method, we follow previous work using cross-attention with FiD to measure the relevance of context passages for question-answering~\cite{DKRR, atlasqa}.
The calculated relevance scores are used to rerank the passages.
Given a query and a group of 100 context passages that might aid in answering the query, \ScoreFiD reranks the context passages in descending order by the relevance scores calculated using the cross-attention scores as the FiD model tries to answer the query.

Figure~\ref{fig:scorefid} shows the setup of \ScoreFiD.
As the FiD model processes each passage separately alongside the query, the input prompt for each query--passage pair starts with \texttt{question:} followed by the query and then \texttt{context:} followed by the passage text.
This input is formed for each passage.
To further refine the calculation of the relevance scores, instead of averaging relevance scores over all tokens in the input, the relevance scores for the tokens corresponding to the text before the passage text are zeroed out.
We found that this helped reranking effectiveness.

\section{Experimental Setup}

\subsection{Datasets}
Following similar work for zero-shot listwise re-ranking~\cite{pradeep2023rankvicuna, pradeep2023rankzephyr, zhang2023rankwithoutgpt}, our evaluation focused on testing zero-shot reranking capabilities across a diverse range of test collections:\
the TREC 2019~\cite{craswell2020overview} and 2020~\cite{craswell2021overview} Deep Learning Tracks from the MS MARCO v1 passage ranking task and the TREC 2021~\cite{craswell2022overview} and 2022~\cite{craswell2023overview} Deep Learning Tracks from the MS MARCO v2 passage ranking task (referred to as DL19--DL22 for simplicity).
In addition, we also test using the BEIR collection~\cite{thakur2021beir}, which spans a variety of diverse text retrieval tasks and domains.
We test on all BEIR tasks except for CQADupStack.

\subsection{Model Training and Hyperparameters}

The models were trained in the T5x library~\cite{t5x} to enable training using TPUs.
Training hyperparameters were chosen to closely align with FiD work from ~\citet{atlasqa}.
In particular, training was done with a batch size of 64, a dropout rate of 10\%, and an AdamW optimizer with a peak learning rate of 0.00005 along a linear warmup for 100 steps.

Following Atlas~\cite{atlasqa}, the learning rate was kept the same for all models, but smaller models were sometimes trained for fewer epochs.
Models were trained for varying epochs depending on the model size and depending on the reranking method being considered.
For more information on how the models were trained differently depending on model size, see Tables \ref{tab:rankFiD_ablation} and \ref{tab:scoreFiD_ablation} in the Ablation Studies section of this work.

The models were initialized with T5 1.1 LM-Adapted weights~\cite{t5} instead of the superior FLAN-T5 weights~\cite{chung2022scaling} to maintain the zero-shot claim of this work as~\citet{pradeep2023rankvicuna} have shown that the FLAN mixture includes the MS MARCO QA task.
Additionally, the FLAN mixture includes many of the datasets from BEIR as well. 
Our models were initialized with the T5- base, large, and XL variants to test different-sized models.

The length of the concatenation of the query and the passage was limited to 150 tokens using the SentencePiece tokenizer of T5 in the case of MS MARCO passage reranking.
The limit was set to 300 tokens for reranking on BEIR datasets to accommodate longer queries and passages.

With 8 TPU-v3 cores, the training of the \RankFiDbase, \RankFiDlarge, and \RankFiDxl models took approximately 2, 2, and 3 hours respectively and the training of the \ScoreFiDbase, \ScoreFiDlarge, and \ScoreFiDxl models took approximately 5, 13, and 38 hours respectively.
Note that in all cases, models were evaluated using one NVIDIA RTX A6000 GPU.

\smallskip \noindent {\bf \RankFiD.} 
The training of \RankFiD is done identically to RankZephyr~\cite{ pradeep2023rankzephyr}.
In the first stage of training, the same ranked lists as in the \rankZephyr and \rankVicuna works formed from \rankgptthreefive orderings of candidate passages retrieved using BM25 with Pyserini~\cite{Pyserini} for queries in the MS MARCO v1 passage ranking training dataset are distilled into the student \RankFiD.
Depending on the model size, this is done for a variable number of training epochs.
The number of epochs is examined further in the Ablation Studies section of this work.
As outlined by~\citet{pradeep2023rankvicuna}, malformed generations are removed from the training data.

In the second stage of training, the same ranked lists as the RankZephyr work from 5k \rankgptfour orderings of candidate passages retrieved with OpenAI's \ada as the first stage from the MS MARCO v1 passage ranking training dataset are distilled into the student \RankFiD for one epoch.
Just as in \rankZephyr, the \RankFiD models are also trained on a subset of $p \; (\leq 20)$ passages, randomly chosen from the
original input to add to the training set for the second stage of training.

\RankFiD models were trained using up to 20 input passages to maintain a window size of 20, matching previous work for listwise reranking using RankGPT or training data from RankGPT~\cite{RankGPT, pradeep2023rankvicuna, pradeep2023rankzephyr}.

\smallskip \noindent {\bf \ScoreFiD.}
The training of the \ScoreFiD models is done without any sort of passage relevance labels, neither from human judgements nor from a teacher model.
Nonetheless, we show it is a strong reranker.
Given questions from the open-domain variants of the Natural Questions~\cite{nq, opennq} and TriviaQA~\cite{Trivia} datasets, the top 100 passages retrieved from the MS MARCO v1 passage corpus~\cite{bajaj2016ms} using BM25 are used as context passages to train a FiD model to generate the corresponding answer to the question.
Unless otherwise specified, models are trained for two epochs.

\ScoreFiD models were trained using 100 input passages as context for question-answering, matching previous using the FiD model for open-domain question-answering~\cite{FiD, DKRR, tamber2023pre}.

\section{Results}

\begin{table*}[t]
\centering \scalebox{0.75}{
\begin{tabular}{lcllllll}
\toprule
\toprule
\multicolumn{1}{l}{\multirow{2}{*}{\hspace{100pt}}} & 
\multicolumn{1}{l}
{\multirow{2}{*}
{\begin{tabular}
[c]{@{}c}\textbf{Model} \\
\textbf{Parameters}
\end{tabular}}} &
\multicolumn{2}{l}
{\multirow{2}{*}
{\begin{tabular}
[c]{@{}c@{}}\textbf{Source} \\
\textbf{Prev.} \hspace{35pt}  \textbf{Top-$k$}
\end{tabular}}} &
\multicolumn{2}{l}
{\multirow{2}{*}
{\begin{tabular}
[c]{@{}c@{}}\textbf{MSv1} \\
\textbf{DL19} \hspace{8pt}  \textbf{DL20}
\end{tabular}}} &
\multicolumn{2}{l}
{\multirow{2}{*}
{\begin{tabular}
[c]{@{}c@{}}\textbf{MSv2} \\
\textbf{DL21} \hspace{8pt}  \textbf{DL22}
\end{tabular}}} \\
\multicolumn{1}{l}{}\ & \multicolumn{1}{l}{}\ & \multicolumn{2}{c}{} & \multicolumn{2}{c}{} \\
\midrule
(1) SPLADE++ ED & 110M & \multicolumn{2}{l}{None}{$|C|$} &
0.731 & 0.720  & 0.684 & 0.570 \\
(2) \repllama & 7B & \multicolumn{2}{l}{None}{$|C|$} &
0.743 & 0.721 & - & - \\
\midrule
(3) \rankllama & 7B &
\multicolumn{2}{l}{\repllama}{100} & 0.756 & 0.774 & - & - \\
\midrule
(4) Rank-wo-GPT & 7B &
\multicolumn{2}{l}{\repllama}{100} & 0.730 & 0.700 & - & - \\
\midrule
(5) \rankVicuna & 7B &
\multicolumn{2}{l}{SPLADE++ ED}{100}            & 
0.746 & 0.747 &   0.701 & 0.582 \\
(6) \rankZephyr & 7B & \multicolumn{2}{l}{SPLADE++ ED}{100} &
0.782  &
0.816  & 0.760 & 0.669 \\
(7) \rankgptfour & ? &
\multicolumn{2}{l}{SPLADE++ ED}{100}                   &
0.746 & 0.708  & 0.772 & 0.718 \\
\midrule
(8a) \RankFiDbase & 220M & \multicolumn{2}{l}{SPLADE++ ED}{100}  & 0.748 & 0.749 &  0.715  & 0.620 \\
(8b) \RankFiDlarge & 770M & \multicolumn{2}{l}{SPLADE++ ED}{100}  & 0.772 & 0.763 &  0.728  &   0.673 \\
(8c) \RankFiDxl & 3B & \multicolumn{2}{l}{SPLADE++ ED}{100}  &  0.771 &  0.768 &   0.746  &  0.687 \\
\bottomrule 
\bottomrule                                            
\end{tabular}}
\caption{nDCG@10 on the DL19 to Dl22 collections. Each reranker uses either the top 100 SPLADE++ ED or the top 100 \repllama retrieved results as input. Note, results for \rankgptfour are taken from~\citet{pradeep2023rankzephyr}.}
\label{tab:main}
\end{table*}

\RankFiD and \ScoreFiD demonstrate effective zero-shot reranking capabilities across all model sizes with \RankFiD showing notable strength.
We compare our methods with previous work across the MS MARCO DL19--22 and BEIR collections using nDCG@10 scores, specifically for reranking the top 100 documents returned by either BM25 first-stage retrieval or SPLADE++ EnsembleDistil (SPLADE++ ED) first-stage retrieval~\cite{formal2022distillation}.
BM25 serves as a basic ``bag-of-words'' baseline, while SPLADE++ ED serves as the stronger supervised first-stage method.

\subsection{MS MARCO Collections}

Table \ref{tab:main} first presents reranking results for the \RankFiD method with SPLADE++ ED first-stage retrieval.
To compare with recent work in LLM rerankers~\cite{ma2023finetuning, zhang2023rankwithoutgpt}, we also include results for  \repllama~\cite{ma2023finetuning} first-stage dense retrieval.

We first observe that \RankFiD's reranking effectiveness, shown in rows 8(a--c), tends to improve with increased model sizes. 
The scores for \RankFiDbase are the lowest for DL19--DL22, while the scores for \RankFiDxl are the highest, except for on DL19 where \RankFiDlarge scores similarly.

Noting first-stage retrieval methods in rows 1 vs 2, \repllama has stronger effectiveness than SPLADE++ ED on DL19 and DL20.
Rows 3 and 4 show the results of the 7B parameter model variants of \rankllama~\cite{ma2023finetuning} and Rank-wo-GPT~\cite{zhang2023rankwithoutgpt} reranking \repllama first-stage retrieval.
Notably, both \rankllama and Rank-wo-GPT are supervised methods, with \rankllama being a pointwise reranker and Rank-wo-GPT being a listwise reranker.
Even the smallest \RankFiD model, \RankFiDbase achieves better reranking results than Rank-wo-GPT.
Further, \RankFiDlarge and \RankFiDxl are competitive with \rankllama, achieving stronger results on DL19, but weaker results on DL20.
Our results are competitive, despite the substantially smaller model sizes, the lack of supervised passage relevance training data, and weaker SPLADE++ ED first-stage retrieval.

Table \ref{tab:main} shows that all \RankFiD models surpass the much larger \rankVicuna model, shown in rows 8(a--c) vs 5.
Comparing to \rankZephyr, we see that the \RankFiD models tend to perform worse, however, \RankFiDlarge and \RankFiDxl have stronger reranking effectiveness than \rankZephyr on DL22, shown in rows \mbox{8(b--c)} vs 6.
Finally, compared to the state-of-the-art \rankgptfour, shown in row 7, rows 8(a--c) show that all \RankFiD models have stronger reranking effectiveness on DL19 and DL20, but weaker effectiveness on DL21 and DL22. 
Interestingly, the same is true for \rankZephyr, which outperforms \rankgptfour on both DL19 and DL20, but it underperforms \rankgptfour on DL21 and DL22.
This may be a result of \RankFiD and \rankZephyr being trained to rerank MS MARCO v1 passages only.

We have shown that even though \rankgptfour is expected to be some orders of magnitude larger than \RankFiD in parameters and \rankZephyr ranges from 2 to 30 times larger than the \RankFiD models in parameters, our results indicate that \RankFiD can achieve reranking effectiveness that rivals that of \rankgptfour and \rankZephyr.

\begin{table*}[t]
\centering
\scalebox{0.68}{
\begin{tabular}{l|c|ccc|ccc|ccc}
\toprule
\toprule
 & \textbf{BM25} & \multicolumn{3}{c|}{\textbf{\ScoreFiD}} & \multicolumn{3}{c|}{\textbf{\RankFiD}} & \textbf{RankVicuna} & \textbf{RankZephyr} & \textbf{Rank-wo-GPT} \\
 \textbf{Model Size (Parameters)} & - & 220M & 770M & 3B & 220M & 770M & 3B & 7B & 7B & 7B \\
 \midrule
Arguana       & 39.7 &    - &    - &    - & 32.1 & 36.7 & - & - & - & - \\
BioASQ        & 52.3 & 52.8 & 52.9 & 54.4 & 56.6 & 56.6 & 57.5 & - & - & - \\
Climate-FEVER & 16.5 &    - &    - &    - & 22.6 & 19.8 & - & - & - & - \\
DBPedia       & 31.8 &    - &    - &    - & 42.8 & 43.5 & 43.7 & - & - & 42.3 \\
FEVER         & 65.1 &    - &    - &    - & 76.8 & 73.9 & - & - & - & - \\
FiQA          & 23.6 & 33.6 & 37.5 & 41.0 & 36.7 & 41.6 & 42.5 & - & - & 35.1 \\
HotpotQA      & 63.3 & 68.3 & 70.6 & 73.4 & 69.6 & 70.9 & - & - & - & - \\
NFCorpus      & 32.2 &    - &    - &    - & 34.3 & 35.4 & 35.9 & - & - & 32.8 \\
NQ            & 30.6 &    - &    - &    - & 52.0 & 55.3 & - & - & - & - \\
Quora         & 78.9 &    - &    - &    - & 83.1 & 79.1 & - & - & - & - \\
Robust04      & 40.7 & 46.7 & 51.8 & 55.8 & 48.7 & 55.3 & 57.2 & - & - & - \\
SCIDOCS       & 14.9 &    - &    - &    - & 16.4 & 18.1 & - & - & - & 16.2 \\
SciFact       & 67.9 &    - &    - &    - & 71.3 & 74.1 & 73.5 & - & - & 64.7 \\
Signal-1M     & 33.0 &    - &    - &    - & 34.3 & 31.0 & 31.0 & - & - & - \\
Touche-2020   & 44.2 &    - &    - &    - & 33.1 & 33.4 & 34.4 & - & - & - \\
TREC-COVID    & 59.5 & 70.3 & 75.1 & 77.8 & 76.8 & 80.3 & 79.0 & 79.8 & 83.8 & 80.4 \\
TREC-NEWS     & 39.5 & 42.4 & 43.6 & 44.0 & 44.5 & 46.7 & 47.6 & 46.7 & 51.8 & - \\

\midrule
\textbf{Average (Overall)}       & 43.2 &    - &    - &    - & 48.9 & 50.1 & - & - & - & - \\
\textbf{Average (QA Retrieval)}  & 46.5 & 52.4 & 55.3 & 57.7 & 55.5 & 58.6 & - & - & - & - \\
\bottomrule
\bottomrule
\end{tabular}
}
\vspace{0.2cm}
\caption{nDCG@10 scores for reranking the top 100 passages returned by BM25 on BEIR.
}
%\vspace{-0.4cm}
\label{tab:beir_bm25}
\end{table*}

\subsection{BEIR}

For BEIR test collections, we examine reranking the top 100 passages retrieved using BM25 first-stage retrieval in Table \ref{tab:beir_bm25}.
Since \ScoreFiD is designed to work for reranking in question-answering contexts, we focus on BEIR datasets that involve retrieval for question answering, excluding the Natural Questions dataset due to our focus on zero-shot reranking and the fact that \ScoreFiD's training involved data from Natural Questions.
The last row of the table shows the average nDCG@10 score for BEIR datasets involving question-answering retrieval and the second-last row shows the average nDCG@10 score overall.
We observe that reranking with both \RankFiD and \ScoreFiD improves nDCG@10 scores compared to BM25 scores, suggesting that these methods generalize well.
The tables show that as the models are scaled up in parameters, the average scores improve.

While the \RankFiDxl model wasn't tested on some BEIR datasets due to time constraints, the data we have show that it typically delivers the strongest results among all our models.
We also observe that \RankFiD has stronger results compared to \ScoreFiD, with higher average scores for question-answering retrieval.

For both \rankVicuna and \rankZephyr, only TREC-COVID and TREC-NEWS results are made available.
Comparing with \rankVicuna, we see that both the \RankFiDlarge and \RankFiDxl models are competitive, with the \RankFiDlarge model performing better on TREC-COVID and performing similarly on TREC-NEWS.
The \RankFiDxl model performs slightly worse on TREC-COVID, but better on TREC-NEWS.
Comparing with \rankZephyr, we observe that like results seen previously with MS MARCO passage reranking, \rankZephyr tends to perform better.

Additionally, we include comparisons with Rank-wo-GPT, which has results available for reranking the top 100 documents retrieved using BM25 on select BEIR datasets.
Despite the authors of Rank-wo-GPT noting its limited generalization capability, the model provides a valuable reference point due to the scarcity of comparable studies in listwise reranking.
Focusing on the 7B model variant from Rank-wo-GPT, our analysis reveals that even the smallest variant of \RankFiD, the \RankFiDbase model, tends to yield stronger results.

\subsection{Model Efficiency}

Our model variants are designed with significantly fewer parameters compared to the recent listwise reranking models \rankVicuna, \rankZephyr, and Rank-wo-GPT.
Typically, larger LLMs are expected to deliver better effectiveness and broader generalizability.
Our approaches vary in size with the different \RankFiD and \ScoreFiD model variants initialized using different size T5 models providing a tradeoff between computation needed and reranking effectiveness.

\begin{figure}[t]
\centering
\hspace*{-0.5cm}
\scalebox{0.35}{
\includegraphics{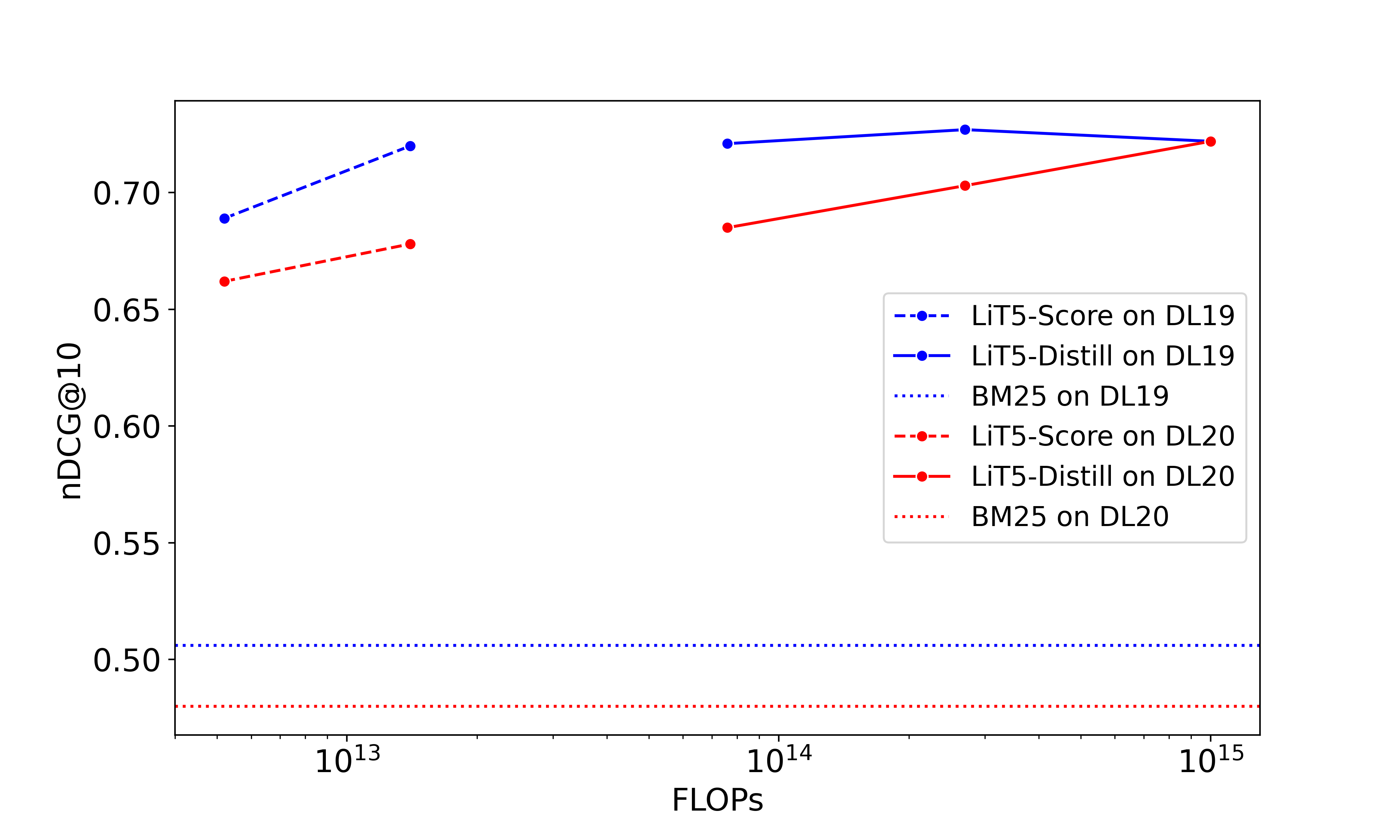}
}
\caption{nDCG@10 on DL19 and DL20 for the \ScoreFiD and \RankFiD model variants reranking 100 BM25 retrieved passages as a function of FLOPs (log-scale) to rerank passages for a single query from DL19.}

\label{fig:flops_figure}
\end{figure}

\begin{table*}[t]
\centering \scalebox{0.75}{
\begin{tabular}{lcllllll}
\toprule
\toprule
\multicolumn{1}{l}{\multirow{2}{*}{\hspace{100pt}}} & 
\multicolumn{1}{l}
{\multirow{2}{*}
{\begin{tabular}
[c]{@{}c}\textbf{Model} \\
\textbf{Parameters}
\end{tabular}}} &
\multicolumn{2}{l}
{\multirow{2}{*}
{\begin{tabular}
[c]{@{}c@{}}\textbf{Source} \\
\textbf{Prev.} \hspace{35pt}  \textbf{Top-$k$}
\end{tabular}}} &
\multicolumn{2}{l}
{\multirow{2}{*}
{\begin{tabular}
[c]{@{}c@{}}\textbf{MSv1} \\
\textbf{DL19} \hspace{8pt}  \textbf{DL20}
\end{tabular}}} &
\multicolumn{2}{l}
{\multirow{2}{*}
{\begin{tabular}
[c]{@{}c@{}}\textbf{MSv2} \\
\textbf{DL21} \hspace{8pt}  \textbf{DL22}
\end{tabular}}} \\
\multicolumn{1}{l}{}\ & \multicolumn{1}{l}{}\ & \multicolumn{2}{c}{} & \multicolumn{2}{c}{} \\
\midrule
(1a) BM25 & - & \multicolumn{2}{l}{None}{$|C|$} &
0.506 & 0.480 &  0.446 & 0.269 \\
\midrule
(2a) MonoT5 & 220M &
\multicolumn{2}{l}{BM25}{100} & 0.715 & 0.670 & - & - \\
(2b) MonoT5 & 3B &
\multicolumn{2}{l}{BM25}{100} & 0.718 & 0.689 & - & - \\
(3a) RankT5 & 3B &
\multicolumn{2}{l}{BM25}{100} & 0.712 & 0.695 & - & - \\
(4a) Rank-wo-GPT & 7B &
\multicolumn{2}{l}{BM25}{100} & 0.718 & 0.674 & - & - \\

\midrule
(5a) PRP-Allpair & 3B &
\multicolumn{2}{l}{BM25}{100}            & 
0.698 & 0.681 & - & - \\

\midrule
(6a) \rankVicuna & 7B &
\multicolumn{2}{l}{BM25}{100}            & 
0.668 & 0.655 &   0.624 & 0.430 \\

(7a) \rankZephyr & 7B & \multicolumn{2}{l}{BM25}{100} &
0.742  & 0.709  & 0.703 & 0.515 \\
(8a) \rankgptthreefive & ? &
\multicolumn{2}{l}{BM25}{100}                   &
0.686 & 0.620  & 0.605 & 0.418 \\
(8b) \rankgptfour & ? &
\multicolumn{2}{l}{BM25}{100}                   &
0.750 & 0.704  & 0.707 & 0.508 \\
\midrule
(9a) \RankFiDbase & 220M & \multicolumn{2}{l}{BM25}{100}  & 0.721 & 0.685 & 0.650  &   0.473 \\
(9b) \RankFiDlarge & 770M & \multicolumn{2}{l}{BM25}{100}  & 0.727 & 0.703 & 0.665  & 0.506 \\
(9c) \RankFiDxl & 3B & \multicolumn{2}{l}{BM25}{100}  & 0.722 & 0.722 & 0.686 & 0.513 \\

\midrule
(10a) \ScoreFiDbase & 220M & \multicolumn{2}{l}{BM25}{100}  &  0.689 &  0.662 &    0.615  &    0.430 \\

(10b) \ScoreFiDlarge & 770M & \multicolumn{2}{l}{BM25}{100}  &  0.720 &  0.678 &    0.674  &    0.475 \\

(10c) \ScoreFiDxl & 3B & \multicolumn{2}{l}{BM25}{100}  &  0.700 &  0.657 &    0.666  &    0.469 \\

\bottomrule 
\bottomrule                                            
\end{tabular}}
\caption{nDCG@10 on the DL19 to Dl22 collections. Each reranker uses the top 100 BM25 retrieved results as input. Note results for MonoT5 and RankT5 are taken from~\citet{qin2023large} and results for RankGPT are taken from~\citet{pradeep2023rankzephyr}.}
\label{tab:second}
\end{table*}

In Figure \ref{fig:flops_figure} we show nDCG@10 on DL19 and DL20 for the \RankFiD and \ScoreFiD model variants as a function of FLOPs to rerank one single example from DL19.
Note, we do not include results from the \ScoreFiDxl model because of the weaker results on DL19 and DL20 compared to the smaller \ScoreFiD models, further analyzed in section \ref{sec:ScoreFiDTrainingAblation}.
The figure shows that although reranking effectiveness tends to improve as models are scaled up in parameters, this comes at the cost of increased computation.
We see that \ScoreFiD models take less computation than \RankFiD models.
This is because \RankFiD's sliding window strategy must repeatedly go over the same passages multiple times to rerank 100 passages.
Additionally, \ScoreFiD only tends to generate a few tokens of output as a result of being trained to generate short answers for questions from the open-domain versions of Natural Questions and TriviaQA~\cite{nq, opennq, Trivia}.
On the other hand, \RankFiD has to generate many more tokens to generate an ordering for a ranked list in the form [] > [], e.g., [4] > [2] > ... > [5] for 20 passages.

\section{Ablation Studies}

\subsection{BM25 First-Stage}

In Table \ref{tab:second}, we focus on reranking MS MARCO passages with BM25 first-stage retrieval.
Rows \mbox{9(a--c)} vs 10(a--c) show that \RankFiD consistently surpasses \ScoreFiD in reranking effectiveness across all model sizes and datasets.

When our methods are compared to those employing the T5 model in supervised settings, even \ScoreFiD, though the less effective of our two methods, is competitive.
This is despite both our methods being zero-shot.
Both MonoT5~\cite{monoT5} and RankT5~\cite{zhuang2022rankt5} are supervised methods that leverage the T5 model to perform reranking.
\RankFiD surpasses MonoT5 and RankT5 across the different datasets and model sizes, shown in rows 2(a) vs 9(a) and rows 2(b) and 3(a) vs 9(c).
Additionally, \ScoreFiD shows strong reranking effectiveness, with the 770M parameter model outperforming all MonoT5 and RankT5 variants on DL19, shown in rows 2(a), 2(b), and 3(a) vs 10(b).

\RankFiD's reranking effectiveness continues to improve with increased model sizes.
However, in the case of \ScoreFiD, while the T5-large variant with 770M parameters outperforms the T5-base variant with 220M parameters, the T5-XL 3B parameter variant does not exhibit as strong reranking effectiveness as the large variant.
Further details on this observation are shared in Section \ref{sec:ScoreFiDTrainingAblation}.

Both \RankFiD and \ScoreFiD show stronger reranking effectiveness compared to \rankgptthreefive and \rankVicuna, which is trained by distilling ranking orderings from \rankgptthreefive.
This is shown in rows 9(a--c) and 10(a--c) vs 6(a) and 8(a).
Remarkably, this superior reranking effectiveness of our models holds even for the smaller model variants, despite being compared to much larger \rankVicuna and \gptthreefive models.

The success of the \RankFiD models is largely attributed to their effective use of listwise reranking orderings distilled not only from \rankgptthreefive but also from \rankgptfour, transferred to smaller models.
This strategy enables \RankFiD, as shown in rows 9(a--c), to outshine both \rankgptthreefive and \rankVicuna.
\RankFiD continues to be competitive with \rankgptfour and \rankZephyr, featured in rows 8(b) and 7(a), respectively.
The largest \RankFiD model, in row 9(c), excels over \rankgptfour in DL20 and DL22 and over \rankZephyr in DL20.
Effectiveness gains from training with \rankgptfour orderings are further discussed in Section \ref{sec:RankFiDTrainingAblation}.

Comparing results with Rank-wo-GPT, we observe that not only do the \RankFiD models continue to show stronger results than the 7B parameter Rank-wo-GPT model variant, shown in rows 9(a--c) vs 4(a), although while depending on GPT models for training, the zero-shot \ScoreFiDlarge model also has stronger results, shown in rows 10(b) vs 4(a), whilst not depending on GPT models.

A recent study by~\citet{qin2023large} argued that LLMs struggle with listwise ranking formulations, often producing inconsistent or irrelevant outputs.
This work made use of T5 models ranging from 3B to 11B parameters instruction fine-tuned with the FLAN mixture~\cite{chung2022scaling} to perform pairwise reranking.
Compared to the best 3B parameter model from this work, shown in row 5(a), all \RankFiD models achieve superior results and \ScoreFiDlarge achieves competitive results.
Notably, our approach achieves this without relying on pre-trained FLAN-T5 model weights and instead using the inferior T5 1.1 LM-Adapted weights to maintain the claim of our methods being zero-shot.

\subsection{\RankFiD Training Ablation} \label{sec:RankFiDTrainingAblation}

\begin{table}[ht!]
\centering
\scalebox{0.73}{%
\begin{tabular}{lcccc}
\toprule
\toprule
\textbf{Model} & \textbf{Model} & \textbf{3.5 Training} & \bf DL19   & \bf DL20  \\
\textbf{Variant} &  \textbf{size} & \textbf{Epochs} & \textbf{nDCG@10} & \textbf{nDCG@10}  \\
\midrule
\multicolumn{5}{l}{RankGPT$_{3.5}$ Distillation $1^{st}$ Stage Training} \\
\arrayrulecolor{gray}\midrule\arrayrulecolor{black}
Base       & 220M & 5   &  0.655  &   0.580  \\
Base       & 220M & 6   &  0.669  &   0.597   \\
Base       & 220M & 7   &  0.659  &   0.577   \\
Large      & 770M & 3   &  0.664  &   0.610   \\
Large      & 770M & 4   &  0.673  &   0.628   \\
Large      & 770M & 5   &  0.676  &   0.619   \\
XL         & 3B   & 1   &  0.664  &   0.611   \\
XL         & 3B   & 2   &  0.664  &   0.626   \\
\midrule
\multicolumn{5}{l}{RankGPT$_{4}$ Distillation $2^{nd}$ Stage Training} \\
\arrayrulecolor{gray}\midrule\arrayrulecolor{black}
Base       & 220M & 5   &  0.716  &  0.661  \\
Base       & 220M & 6   &  0.710  &  0.666   \\
Large      & 770M & 3   &  0.715  &  0.697   \\
Large      & 770M & 4   &  0.726  &  0.695   \\
XL         & 3B   & 1   &  0.722  &  0.722   \\
XL         & 3B   & 2   &  0.718  &  0.708   \\
\midrule
\multicolumn{5}{l}{RankGPT$_{4}$ Distillation $2^{nd}$ Stage Training, with subset sampling} \\
\arrayrulecolor{gray}\midrule\arrayrulecolor{black}
Base       & 220M & 5   &   0.718  &   0.663  \\
Base       & 220M & 6   &  0.721  &   0.685   \\
Large       & 770M & 3   &  0.727  &   0.703   \\
Large       & 770M & 4   &  0.712  &   0.686   \\
XL          & 3B   & 1   &  0.715  &  0.719   \\
\bottomrule
\bottomrule
\end{tabular}%
}

\caption{The effectiveness of \RankFiD for reranking 100 BM25 retrieved passages as the number of training epochs on RankGPT$_{3.5}$ orderings are varied and training data from RankGPT$_4$ orderings is used.
}
\label{tab:rankFiD_ablation}
\end{table}

Table \ref{tab:rankFiD_ablation} shows the reranking effectiveness of \RankFiD as the number of first-stage training epochs with \rankgptthreefive orderings are varied.
The table shows that for each model, we must be careful not to overfit the models with over-training.

For \RankFiDbase, the best model throughout the first- and second-stages of training is one that is trained for 6 epochs in the first stage.
For \RankFiDlarge, although in the first stage the models trained for 4 and 5 epochs perform strongest, into the second stage when the models are trained to rerank the 3 randomly sampled subsets of $p \; (\leq 20)$ passages, we see that the \RankFiDlarge model trained for 3 epochs in the first stage has the strongest reranking effectiveness into the second stage.
For \RankFiDxl, in the first stage, the model trained for 2 epochs has stronger reranking effectiveness than the one trained for 1 epoch.
Into the second stage, the model trained for 1 epoch has stronger results, but this model has worse results when we also include training with the 3 sampled subsets as the model likely starts to overfit.
So, the best \RankFiDxl model is the one trained for one epoch in the first stage that does not train using the sampled subsets in the second stage.

\subsection{\ScoreFiD Training Ablation}
\label{sec:ScoreFiDTrainingAblation}

We observe that the largest \ScoreFiD model, \ScoreFiDxl---with 3B parameters---performs worse on all the MS MARCO collections studied compared to the \ScoreFiDlarge model and compared to the \ScoreFiDbase model on DL20.
Nonetheless, the \ScoreFiDxl model had stronger reranking effectiveness than \ScoreFiDbase and \ScoreFiDlarge on the BEIR test collections shown in Table \ref{tab:beir_bm25}.

\ScoreFiDxl is trained for only 1 epoch.
Table \ref{tab:scoreFiD_ablation} shows that \ScoreFiDxl sees an effectiveness drop on DL19 and DL20 when training for more than 1 epoch, while \ScoreFiDbase and \ScoreFiDlarge have strong reranking effectiveness with 2 epochs of training.
Perhaps, \ScoreFiDxl has reduced effectiveness because it is more prone to overfitting.
This is a curious result that should be examined further in future work to help scale the \ScoreFiD method to larger models.

\begin{table}[t]

\centering
\scalebox{0.8}{%
\begin{tabular}{lccc}
\toprule
\toprule
\textbf{Model} & \textbf{Training} & \bf DL19   & \bf DL20  \\
\textbf{Variant} & \textbf{Epochs} & \textbf{nDCG@10} & \textbf{nDCG@10}  \\
\midrule
Base      & 2 &  0.689  &   0.662  \\
Large     & 2 &  0.720  &   0.678   \\
XL        & 1 &  0.700  &   0.657   \\
XL        & 2 &  0.676  &   0.645   \\
\bottomrule
\bottomrule
\end{tabular}%
}
\caption{The effectiveness of \ScoreFiD for reranking 100 BM25 retrieved passages as the number of training epochs for \ScoreFiDxl is varied.
}
\label{tab:scoreFiD_ablation}
\end{table}

\section{Conclusion and Future Work}

In this work, we introduce two families of zero-shot listwise rerankers using encoder--decoder models, \RankFiD and \ScoreFiD.

\RankFiD shows particularly strong reranking results.
We show that listwise reranking effectiveness from \rankgptthreefive and \rankgptfour can be distilled into small encoder--decoder models, while still maintaining strong reranking effectiveness.
Our largest \RankFiD model, \RankFiDxl, in some cases, outperforms the current state-of-the-art listwise rerankers \rankgptfour and \rankZephyr, despite having fewer parameters and building on outdated T5 model weights.

\ScoreFiD followed previous work in using relevance scores calculated from cross-attention scores.
We show that these relevance scores are a good measure of passage importance to queries and allow for strong zero-shot listwise rerankers for question-answering retrieval tasks.
\ScoreFiD also shows that listwise reranking can be done in LLMs other than the GPT models without the need for training using passage relevance labels from human judgements or from rankings from teacher models while still having strong effectiveness.

This work presents the smallest models that we know of that perform zero-shot listwise reranking with LLMs.
\RankFiD and \ScoreFiD are competitive with both current work in zero-shot reranking using modern advances in LLMs and supervised methods using the same T5 models as \RankFiD and \ScoreFiD.
We test models ranging from 220M to 3B parameters for both the \RankFiD and \ScoreFiD families showing that small 220M parameter models can still achieve strong results for listwise reranking, all while being able to generalize well to out-of-domain reranking tasks and being relatively quick to train and run.

The findings from this work have some interesting applications for future work.
We show that we can successfully distill reranking effectiveness from large GPT models to much smaller models in a retrieval-augmented generation setting.
Future work may find it interesting to consider what other LLM behavior can be distilled down to smaller models when working with input passages.
We also show that relevance scores calculated using cross-attention scores can be strong measures of the importance of input passages in text generation.
An application of this in future work may be to generate references or citations for the output of a retrieval-augmented generation model without needing to explicitly train for the generation of citations.

\section*{Acknowledgements}

This research was supported in part by the Natural Sciences and Engineering Research Council (NSERC) of Canada.
We thank Google Cloud and the TPU Research Cloud Program for credits to support some of our experimental runs.

\bibliography{custom}

\begin{thebibliography}{30}
\expandafter\ifx\csname natexlab\endcsname\relax\def\natexlab#1{#1}\fi

\bibitem[{Bajaj et~al.(2016)Bajaj, Campos, Craswell, Deng, Gao, Liu, Majumder, McNamara, Mitra, Nguyen, Rosenberg, Song, Stoica, Tiwary, and Wang}]{bajaj2016ms}
Payal Bajaj, Daniel Campos, Nick Craswell, Li~Deng, Jianfeng Gao, Xiaodong Liu, Rangan Majumder, Andrew McNamara, Bhaskar Mitra, Tri Nguyen, Mir Rosenberg, Xia Song, Alina Stoica, Saurabh Tiwary, and Tong Wang. 2016.
\newblock {MS MARCO: A Human Generated MAchine Reading COmprehension Dataset}.
\newblock \emph{arXiv:1611.09268v3}.

\bibitem[{Child et~al.(2019)Child, Gray, Radford, and Sutskever}]{child2019generating}
Rewon Child, Scott Gray, Alec Radford, and Ilya Sutskever. 2019.
\newblock {Generating Long Sequences with Sparse Transformers}.
\newblock \emph{arXiv:1904.10509}.

\bibitem[{Chung et~al.(2022)Chung, Hou, Longpre, Zoph, Tay, Fedus, Li, Wang, Dehghani, Brahma, Webson, Gu, Dai, Suzgun, Chen, Chowdhery, Castro-Ros, Pellat, Robinson, Valter, Narang, Mishra, Yu, Zhao, Huang, Dai, Yu, Petrov, Chi, Dean, Devlin, Roberts, Zhou, Le, and Wei}]{chung2022scaling}
Hyung~Won Chung, Le~Hou, Shayne Longpre, Barret Zoph, Yi~Tay, William Fedus, Yunxuan Li, Xuezhi Wang, Mostafa Dehghani, Siddhartha Brahma, Albert Webson, Shixiang~Shane Gu, Zhuyun Dai, Mirac Suzgun, Xinyun Chen, Aakanksha Chowdhery, Alex Castro-Ros, Marie Pellat, Kevin Robinson, Dasha Valter, Sharan Narang, Gaurav Mishra, Adams Yu, Vincent Zhao, Yanping Huang, Andrew Dai, Hongkun Yu, Slav Petrov, Ed~H. Chi, Jeff Dean, Jacob Devlin, Adam Roberts, Denny Zhou, Quoc~V. Le, and Jason Wei. 2022.
\newblock {Scaling Instruction-Finetuned Language Models}.
\newblock \emph{arXiv:2210.11416}.

\bibitem[{Craswell et~al.(2020)Craswell, Mitra, Yilmaz, and Campos}]{craswell2021overview}
Nick Craswell, Bhaskar Mitra, Emine Yilmaz, and Daniel Campos. 2020.
\newblock Overview of the {TREC} 2020 deep learning track.
\newblock In \emph{Proceedings of the Twenty-Ninth Text REtrieval Conference Proceedings (TREC 2020)}, Gaithersburg, Maryland.

\bibitem[{Craswell et~al.(2021)Craswell, Mitra, Yilmaz, Campos, and Lin}]{craswell2022overview}
Nick Craswell, Bhaskar Mitra, Emine Yilmaz, Daniel Campos, and Jimmy Lin. 2021.
\newblock Overview of the {TREC} 2021 deep learning track.
\newblock In \emph{Proceedings of the Thirtieth Text REtrieval Conference (TREC 2021)}.

\bibitem[{Craswell et~al.(2022)Craswell, Mitra, Yilmaz, Campos, Lin, Voorhees, and Soboroff}]{craswell2023overview}
Nick Craswell, Bhaskar Mitra, Emine Yilmaz, Daniel Campos, Jimmy Lin, Ellen~M. Voorhees, and Ian Soboroff. 2022.
\newblock Overview of the {TREC} 2022 deep learning track.
\newblock In \emph{Proceedings of the Thirty-First Text REtrieval Conference (TREC 2021)}, Gaithersburg, Maryland.

\bibitem[{Craswell et~al.(2019)Craswell, Mitra, Yilmaz, Campos, and Voorhees}]{craswell2020overview}
Nick Craswell, Bhaskar Mitra, Emine Yilmaz, Daniel Campos, and Ellen~M. Voorhees. 2019.
\newblock Overview of the {TREC} 2019 deep learning track.
\newblock In \emph{Proceedings of the Twenty-Eighth Text REtrieval Conference Proceedings (TREC 2019)}, Gaithersburg, Maryland.

\bibitem[{de~Jong et~al.(2022)de~Jong, Zemlyanskiy, Ainslie, FitzGerald, Sanghai, Sha, and Cohen}]{de2022fido}
Michiel de~Jong, Yury Zemlyanskiy, Joshua Ainslie, Nicholas FitzGerald, Sumit Sanghai, Fei Sha, and William Cohen. 2022.
\newblock {FiDO: Fusion-in-Decoder optimized for stronger performance and faster inference}.
\newblock \emph{arXiv:2212.08153}.

\bibitem[{Formal et~al.(2022)Formal, Lassance, Piwowarski, and Clinchant}]{formal2022distillation}
Thibault Formal, Carlos Lassance, Benjamin Piwowarski, and St\'{e}phane Clinchant. 2022.
\newblock {From Distillation to Hard Negative Sampling: Making Sparse Neural IR Models More Effective}.
\newblock In \emph{Proceedings of the 45th International ACM SIGIR Conference on Research and Development in Information Retrieval}, SIGIR '22, page 2353–2359, New York, NY, USA. Association for Computing Machinery.

\bibitem[{Izacard and Grave(2021{\natexlab{a}})}]{DKRR}
Gautier Izacard and Edouard Grave. 2021{\natexlab{a}}.
\newblock {Distilling Knowledge from Reader to Retriever for Question Answering}.
\newblock In \emph{ICLR 2021-9th International Conference on Learning Representations}.

\bibitem[{Izacard and Grave(2021{\natexlab{b}})}]{FiD}
Gautier Izacard and Edouard Grave. 2021{\natexlab{b}}.
\newblock {Leveraging Passage Retrieval with Generative Models for Open Domain Question Answering}.
\newblock In \emph{EACL 2021-16th Conference of the European Chapter of the Association for Computational Linguistics}, pages 874--880. Association for Computational Linguistics.

\bibitem[{Izacard et~al.(2022)Izacard, Lewis, Lomeli, Hosseini, Petroni, Schick, Dwivedi-Yu, Joulin, Riedel, and Grave}]{atlasqa}
Gautier Izacard, Patrick Lewis, Maria Lomeli, Lucas Hosseini, Fabio Petroni, Timo Schick, Jane Dwivedi-Yu, Armand Joulin, Sebastian Riedel, and Edouard Grave. 2022.
\newblock {Few-shot Learning with Retrieval Augmented Language Models}.
\newblock \emph{arXiv:2208.03299}.

\bibitem[{Joshi et~al.(2017)Joshi, Choi, Weld, and Zettlemoyer}]{Trivia}
Mandar Joshi, Eunsol Choi, Daniel~S. Weld, and Luke Zettlemoyer. 2017.
\newblock {TriviaQA: A Large Scale Distantly Supervised Challenge Dataset for Reading Comprehension}.
\newblock In \emph{Proceedings of the 55th Annual Meeting of the Association for Computational Linguistics (Volume 1: Long Papers)}, pages 1601--1611, Vancouver, Canada.

\bibitem[{Kwiatkowski et~al.(2019)Kwiatkowski, Palomaki, Redfield, Collins, Parikh, Alberti, Epstein, Polosukhin, Devlin, Lee, Toutanova, Jones, Kelcey, Chang, Dai, Uszkoreit, Le, and Petrov}]{nq}
Tom Kwiatkowski, Jennimaria Palomaki, Olivia Redfield, Michael Collins, Ankur Parikh, Chris Alberti, Danielle Epstein, Illia Polosukhin, Jacob Devlin, Kenton Lee, Kristina Toutanova, Llion Jones, Matthew Kelcey, Ming-Wei Chang, Andrew~M. Dai, Jakob Uszkoreit, Quoc Le, and Slav Petrov. 2019.
\newblock {Natural Questions: A Benchmark for Question Answering Research}.
\newblock \emph{Transactions of the Association for Computational Linguistics}, 7:452--466.

\bibitem[{Lee et~al.(2019)Lee, Chang, and Toutanova}]{opennq}
Kenton Lee, Ming-Wei Chang, and Kristina Toutanova. 2019.
\newblock {Latent Retrieval for Weakly Supervised Open Domain Question Answering}.
\newblock In \emph{Proceedings of the 57th Annual Meeting of the Association for Computational Linguistics}, pages 6086--6096.

\bibitem[{Lin et~al.(2021)Lin, Ma, Lin, Yang, Pradeep, and Nogueira}]{Pyserini}
Jimmy Lin, Xueguang Ma, Sheng-Chieh Lin, Jheng-Hong Yang, Ronak Pradeep, and Rodrigo Nogueira. 2021.
\newblock {Pyserini}: A {Python} toolkit for reproducible information retrieval research with sparse and dense representations.
\newblock In \emph{Proceedings of the 44th Annual International ACM SIGIR Conference on Research and Development in Information Retrieval (SIGIR 2021)}, pages 2356--2362.

\bibitem[{Ma et~al.(2023{\natexlab{a}})Ma, Wang, Yang, Wei, and Lin}]{ma2023finetuning}
Xueguang Ma, Liang Wang, Nan Yang, Furu Wei, and Jimmy Lin. 2023{\natexlab{a}}.
\newblock {Fine-Tuning LLaMA for Multi-Stage Text Retrieval}.
\newblock \emph{arXiv:2310.08319}.

\bibitem[{Ma et~al.(2023{\natexlab{b}})Ma, Zhang, Pradeep, and Lin}]{ma2023zeroshot}
Xueguang Ma, Xinyu Zhang, Ronak Pradeep, and Jimmy Lin. 2023{\natexlab{b}}.
\newblock {Zero-Shot Listwise Document Reranking with a Large Language Model}.
\newblock \emph{arXiv:2305.02156}.

\bibitem[{Mao et~al.(2021)Mao, He, Liu, Shen, Gao, Han, and Chen}]{GAR}
Yuning Mao, Pengcheng He, Xiaodong Liu, Yelong Shen, Jianfeng Gao, Jiawei Han, and Weizhu Chen. 2021.
\newblock {Generation-Augmented Retrieval for Open-Domain Question Answering}.
\newblock In \emph{Proceedings of the 59th Annual Meeting of the Association for Computational Linguistics and the 11th International Joint Conference on Natural Language Processing (Volume 1: Long Papers)}, pages 4089--4100, Online.

\bibitem[{Nogueira et~al.(2020)Nogueira, Jiang, Pradeep, and Lin}]{monoT5}
Rodrigo Nogueira, Zhiying Jiang, Ronak Pradeep, and Jimmy Lin. 2020.
\newblock {Document Ranking with a Pretrained Sequence-to-Sequence Model}.
\newblock In \emph{Findings of the Association for Computational Linguistics: EMNLP 2020}, pages 708--718, Online.

\bibitem[{Pradeep et~al.(2023{\natexlab{a}})Pradeep, Sharifymoghaddam, and Lin}]{pradeep2023rankvicuna}
Ronak Pradeep, Sahel Sharifymoghaddam, and Jimmy Lin. 2023{\natexlab{a}}.
\newblock {RankVicuna: Zero-Shot Listwise Document Reranking with Open-Source Large Language Models}.
\newblock \emph{arXiv:2309.15088}.

\bibitem[{Pradeep et~al.(2023{\natexlab{b}})Pradeep, Sharifymoghaddam, and Lin}]{pradeep2023rankzephyr}
Ronak Pradeep, Sahel Sharifymoghaddam, and Jimmy Lin. 2023{\natexlab{b}}.
\newblock {RankZephyr: Effective and Robust Zero-Shot Listwise Reranking is a Breeze!}
\newblock \emph{arXiv:2312.02724}.

\bibitem[{Qin et~al.(2023)Qin, Jagerman, Hui, Zhuang, Wu, Shen, Liu, Liu, Metzler, Wang et~al.}]{qin2023large}
Zhen Qin, Rolf Jagerman, Kai Hui, Honglei Zhuang, Junru Wu, Jiaming Shen, Tianqi Liu, Jialu Liu, Donald Metzler, Xuanhui Wang, et~al. 2023.
\newblock {Large Language Models are Effective Text Rankers with Pairwise Ranking Prompting}.
\newblock \emph{arXiv:2306.17563}.

\bibitem[{Raffel et~al.(2020)Raffel, Shazeer, Roberts, Lee, Narang, Matena, Zhou, Li, and Liu}]{t5}
Colin Raffel, Noam Shazeer, Adam Roberts, Katherine Lee, Sharan Narang, Michael Matena, Yanqi Zhou, Wei Li, and Peter~J. Liu. 2020.
\newblock {Exploring the Limits of Transfer Learning with a Unified Text-to-Text Transformer}.
\newblock \emph{The Journal of Machine Learning Research}, 21(1):5485--5551.

\bibitem[{Roberts et~al.(2023)Roberts, Chung, Mishra, Levskaya, Bradbury, Andor, Narang, Lester, Gaffney, Mohiuddin et~al.}]{t5x}
Adam Roberts, Hyung~Won Chung, Gaurav Mishra, Anselm Levskaya, James Bradbury, Daniel Andor, Sharan Narang, Brian Lester, Colin Gaffney, Afroz Mohiuddin, et~al. 2023.
\newblock {Scaling Up Models and Data with $\texttt{t5x} $ and $\texttt{seqio}$}.
\newblock \emph{Journal of Machine Learning Research}, 24(377):1--8.

\bibitem[{Sun et~al.(2023)Sun, Yan, Ma, Wang, Ren, Chen, Yin, and Ren}]{RankGPT}
Weiwei Sun, Lingyong Yan, Xinyu Ma, Shuaiqiang Wang, Pengjie Ren, Zhumin Chen, Dawei Yin, and Zhaochun Ren. 2023.
\newblock {Is {C}hat{GPT} Good at Search? Investigating Large Language Models as Re-Ranking Agents}.
\newblock In \emph{Proceedings of the 2023 Conference on Empirical Methods in Natural Language Processing}, pages 14918--14937, Singapore.

\bibitem[{Tamber et~al.(2023)Tamber, Pradeep, and Lin}]{tamber2023pre}
Manveer~Singh Tamber, Ronak Pradeep, and Jimmy Lin. 2023.
\newblock {Pre-processing Matters! Improved Wikipedia Corpora for Open-Domain Question Answering}.
\newblock In \emph{European Conference on Information Retrieval}, pages 163--176.

\bibitem[{Thakur et~al.(2021)Thakur, Reimers, R{\"u}ckl{\'e}, Srivastava, and Gurevych}]{thakur2021beir}
Nandan Thakur, Nils Reimers, Andreas R{\"u}ckl{\'e}, Abhishek Srivastava, and Iryna Gurevych. 2021.
\newblock {BEIR: A Heterogenous Benchmark for Zero-shot Evaluation of Information Retrieval Models}.
\newblock \emph{arXiv:2104.08663}.

\bibitem[{Zhang et~al.(2023)Zhang, Hofst{\"a}tter, Lewis, Tang, and Lin}]{zhang2023rankwithoutgpt}
Xinyu Zhang, Sebastian Hofst{\"a}tter, Patrick Lewis, Raphael Tang, and Jimmy Lin. 2023.
\newblock {Rank-without-GPT: Building GPT-Independent Listwise Rerankers on Open-Source Large Language Models}.
\newblock \emph{arXiv:2312.02969}.

\bibitem[{Zhuang et~al.(2023)Zhuang, Qin, Jagerman, Hui, Ma, Lu, Ni, Wang, and Bendersky}]{zhuang2022rankt5}
Honglei Zhuang, Zhen Qin, Rolf Jagerman, Kai Hui, Ji~Ma, Jing Lu, Jianmo Ni, Xuanhui Wang, and Michael Bendersky. 2023.
\newblock {RankT5: Fine-Tuning T5 for Text Ranking with Ranking Losses}.
\newblock In \emph{Proceedings of the 46th International ACM SIGIR Conference on Research and Development in Information Retrieval}, pages 2308--2313.

\end{thebibliography}

\clearpage
\appendix

\begin{figure}[t]
\centering
%\hspace*{-0.8cm}
\scalebox{0.6}{
\includegraphics{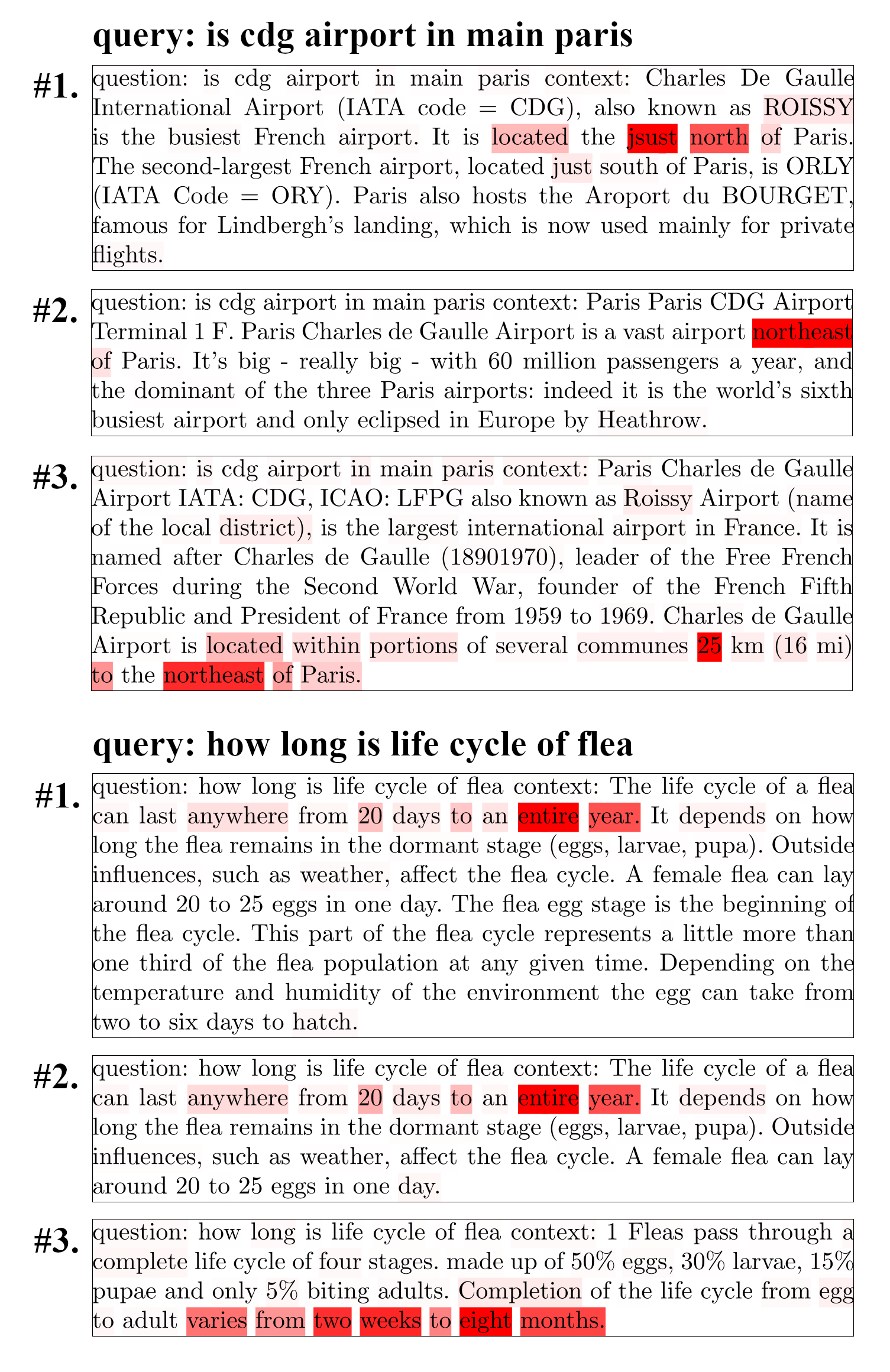}
}
\caption{Visualization of token relevance scores from \ScoreFiDlarge across words of the top 3 reranked MS MARCO v1 passages retrieved using BM25 for two example DL19 queries.}

\label{fig:heatmap}

\end{figure}

\section{\ScoreFiD Relevance Scores}

Figure \ref{fig:heatmap} shows a visualization of the token relevance scores calculated using the cross-attention scores aggregated across words.
Our work shows that overall, these scores are a good way to measure passage relevance to queries, even in a zero-shot setting.
These visualizations provide an interesting look into the inner workings of the \ScoreFiD model and how scores are assigned for passages.
Words with higher scores are highlighted more brightly.

The top 3 passages with the highest average relevance scores are shown for 2 different queries in the DL19 collection.
In both cases, we see that words that could form an answer to the query are highlighted brightly in the passages.
For the query ``is cdg airport in main Paris'', words that suggest the airport is not in Paris are highlighted.
This includes the words ``just'' and ``north'' in the first passage, ``northeast'' in the second and third passages, and the ``25'' corresponding to 25 km northeast to Paris in the third passage.
For the query ``how long is life cycle of flea'', we see a similar pattern.
The second passage is a substring of the first passage and in both cases, the model seems to be highlighting much of the phrase ``anywhere from 20 days to an entire year''.
This is a reasonable answer to the query.
Similarly in the third passage, words from the phrase ``varies from two weeks to eight months'' are highlighted.
The FiD model can recognize an answer to the query within the passage text and this is reflected in the relevance scores calculated per token.

\end{document}